\documentclass[showpacs,twocolumn,aps,prl,superscriptaddress]{revtex4}
\usepackage{amssymb}
\usepackage{amsmath}
\usepackage{epsfig}
\begin{document}

\title{Analogue Hawking Radiation in a dc-SQUID Array  Transmission Line}

\author{P. D. Nation}
\email{paul.d.nation@dartmouth.edu}
\author{M. P. Blencowe}
\author{A. J. Rimberg}
\affiliation{Department of Physics and Astronomy, Dartmouth College, Hanover, New Hampshire 03755, USA}
\author{E. Buks}
\affiliation{Department of Electrical Engineering, Technion, Haifa 32000 Israel}
\date{\today}

\begin{abstract}
We propose the use of a superconducting transmission line formed from an array of dc-SQUID's for investigating analogue Hawking radiation.  Biasing the array with a space-time varying flux modifies the propagation velocity of the transmission line, leading to an effective metric with an horizon.  Being a fundamentally quantum mechanical device, this setup allows for investigations of quantum effects such as back-reaction and analogue space-time fluctuations on the Hawking process. 
\end{abstract}

\pacs{85.25.Dq; 04.80.Cc; 04.70.Dy; 84.40.Az}

\maketitle

\textit{Introduction}.--- The possibility of observing Hawking radiation \cite{hawking:1974} in a condensed matter system was first suggested by Unruh who uncovered the analogy between sound waves in a fluid and a scalar field in curved space-time \cite{unruh:1981}.  In particular, the fluid equations of motion can formally be expressed in terms of an effective metric matching that of a gravitating spherical, non-rotating massive body in Painlev\'e-Gullstrand coordinates \cite{painleve:1921}
\begin{equation}\label{eq:painleve}
ds^{2}=-\left[c_{s}^{2}-v(r)^{2}\right]dt^{2}+2v(r)drdt+dr^{2}+r^{2}d\Omega^{2},
\end{equation}
where $c_{s}$ is the speed of sound and $v(r)$ is the spatially varying velocity of the fluid.  For a sound wave excitation in the fluid, with velocity $c_{s}$, the horizon occurs where $v^{2}(r)=c_{s}^{2}$ and the excitation is incapable of surmounting the fluid flow.   Since Unruh's original proposal, Hawking radiation analogues have been proposed using Bose-Einstein condensates \cite{garay:2000}, liquid Helium \cite{volovik:1999}, electromagnetic transmission lines \cite{schutzhold:2005}, and fiber-optic setups \cite{philbin:2008}.  Estimated Hawking temperatures in these systems vary from a few nano-Kelvin to $10^{3}\mathrm{K}$ respectively, far above temperatures predicted for astronomical black holes and thus usher in the possibility of experimental observation.  Additionally, the understanding of the physics associated with laboratory system analogues may provide clues as to resolving unanswered questions associated with Hawking's original calculation such as the trans-Planckian problem \cite{jacobson:1991}.

\begin{figure}[htbp]
\begin{center}
\includegraphics[width=3.1in]{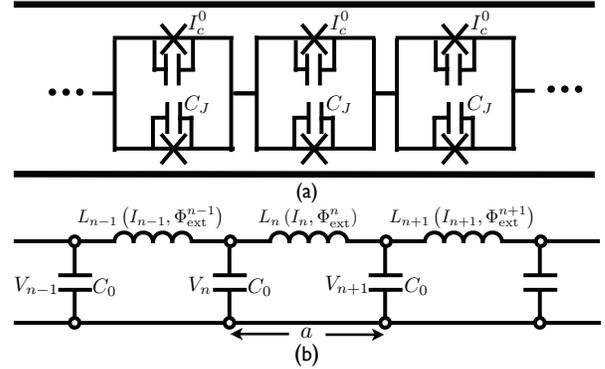}
\caption{a) Layout of the dc-SQUID transmission line. We assume each SQUID element is formed from identical tunnel junctions with critical current $I_{c}$ and capacitance $C_{J}$. b) Effective lumped circuit model valid for frequencies below the plasma frequency and negligible SQUID self-inductance.}
\label{fig:setup}
\end{center}
\end{figure}

In this letter, we propose using a metamaterial formed from an array of direct-current superconducting quantum interference devices (dc-SQUID's).  Modulation of the propagation velocity, necessary for the formation of an horizon, is accomplished through application of an external flux bias through the SQUID loops as indicated in Fig.~\ref{fig:setup}a.
Under appropriate conditions, this configuration provides the superconducting realization of Ref. \cite{schutzhold:2005}, with the benefit of available fabrication methods.  Indeed, arrays of SQUID's with parameters near those required to observe the Hawking effect have already been constructed \cite{beltran:2007,beltran:2008}.  Furthermore, as a quantum device, the SQUID array  goes beyond the capabilities of previously proposed systems, allowing the possibility to probe the effect on Hawking radiation of  quantum fluctuations in the  space-time metric. Thus, in principle, this setup enables the exploration of analogue quantum gravitational effects.

\textit{Model}.--- We consider a coplanar transmission line composed of a centerline conductor formed by a long, $N\gg1$, series array of dc-SQUID's indicated in Fig.~\ref{fig:setup}a.  For simplicity, we assume that all Josephson junctions (JJ) have identical critical current $I_{c}$ and capacitance $C_{J}$ values.  For an individual dc-SQUID, with $\phi_{1}$ and $\phi_{2}$ representing the gauge invariant phases across the JJ's, the equations of motion for $\gamma_{\pm}=\left(\phi_{1}\pm\phi_{2}\right)/2$ take the form
\begin{align}
\frac{1}{\omega_{p}^{2}}\frac{d^{2}\gamma_{+}}{dt^{2}}+\frac{1}{\omega_{c}}\frac{d\gamma_{+}}{dt}+\cos(\gamma_{-})\sin(\gamma_{+})&=\frac{I}{2I_{c}} \notag \\
\frac{1}{\omega_{p}^{2}}\frac{d^{2}\gamma_{-}}{dt^{2}}+\frac{1}{\omega_{c}}\frac{d\gamma_{-}}{dt}+\cos(\gamma_{+})\sin(\gamma_{-})+\frac{2\gamma_{-}}{\beta_{L}}&=\frac{1}{\beta_{L}}\frac{2\pi\Phi_{\mathrm{ext}}}{\Phi_{0}},
\end{align}
with plasma frequency $\omega_{p}=(2\pi I_{c}/C_{J}\Phi_{0})^{1/2}$, characteristic frequency $\omega_{c}=2\pi I_{c}R_{N}/\Phi_{0}$, and normalized self-inductance $\beta_{L}=2\pi L I_{c}/\Phi_{0}$.  The parallel, normal current resistance of the junction is denoted $R_{N}$, while $\Phi_{0}=h/2e$ is the flux quantum and $\Phi_{\mathrm{ext}}$ is the external flux through the SQUID loop.    If $\beta_{L}\ll1$ then the SQUID dynamics can be approximated by a JJ with a flux-tunable critical current, $I^{s}_{c}=2I_{c}\cos(\pi\Phi_{\mathrm{ext}}/\Phi_{0})$, the dynamics of which can be written
\begin{equation}
\frac{1}{(\omega^{s}_{p})^{2}}\frac{d^{2}\gamma_{+}}{dt^{2}}+\sin\left(\gamma_{+}\right)=\frac{I}{I_{c}^{s}},
\end{equation}
where we have dropped the damping term, assuming the  temperature is well below the superconducting critical temperature, and
where the effective plasma frequency is given by $\omega^{s}_{p}=\sqrt{2\pi I^{s}_{c}/(2C_{J}\Phi_{0})}$. We will assume the validity of this approximation and consider a flux-tunable array of Josephson Junctions (JJA).  If we additionally restrict ourselves to frequencies well below the plasma frequency and currents below the critical current, then a JJ behaves as a passive, flux and current dependent inductance given by
\begin{equation}\label{eq:inductance}
L_{n}(I_{n},\Phi^{n}_{\mathrm{ext}})=\frac{\Phi_{0}}{2\pi}\frac{\arcsin\left(I_{n}/I^{s}_{c}\right)}{\left(I_{n}/I^{s}_{c}\right)}
\end{equation}
for the \textit{n}th JJ in the array. The equivalent circuit is given in Fig.~\ref{fig:setup}b where we have labeled the length and capacitance to ground of each JJ by $a$ and $C_{0}$, respectively.  Using Kirchoff's laws, we can write the discrete equations of motion as
\begin{equation}\label{eq:discrete}
V_{n+1}-V_{n}=-\frac{d L_{n}I_{n}}{dt}\ \ ;\ \ I_{n+1}-I_{n}=-C_{0}\frac{d V_{n+1}}{dt}.
\end{equation}
 From (\ref{eq:inductance}), we see that by controlling the external flux bias, or by creating a varying current in the transmission line, we are able to modify the inductance and thus propagation velocity inside the transmission line.  Here, we focus on using the flux degree of freedom as our tunable parameter.  Creating a space-time varying current pulse, as in Ref.~\cite{philbin:2008}, can also be accomplished in our device.  However, our simplified model does not admit the correct dispersion relation to support the required stable nonlinear solitonic localized pulses in the parameter region of interest.  Charge solitons can however be produced in the high impedance regime of our device \cite{haviland:1996}.

\begin{figure}[htbp]
\begin{center}
\includegraphics[width=3.1in]{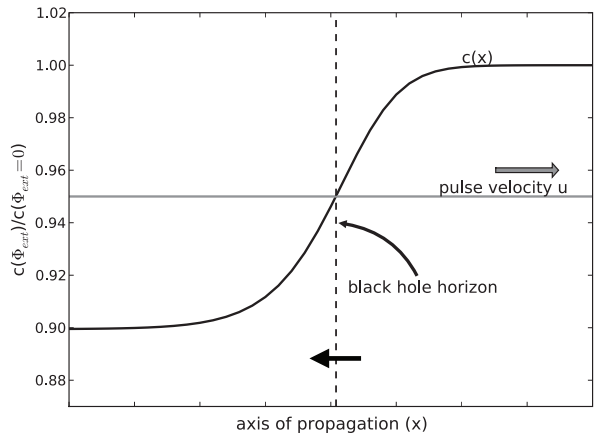}
\caption{Effect of a step-like flux pulse on the propagation velocity of a JJA as seen in the comoving frame.  The pulse velocity was chosen to be $u=0.95c\left(\Phi_{\mathrm{ext}}=0\right)$.  The black hole horizon occurs where $c(x)=u$.  Arrow indicates the only permissible direction of travel across the horizon.}
\label{fig:regions}
\end{center}
\end{figure}

\textit{Effective Geometry and Hawking Temperature}.--- By defining potentials $A_{n}$ such that $I_{n}=-C_{0}dA_{n}/dt$ and $V_{n}=A_{n}-A_{n-1}$ \cite{schutzhold:2005}, the equations of motion (\ref{eq:discrete}) can be combined to yield the discretized wave equation, 
\begin{equation}
\frac{d}{dt}L_{n}C_{0}\frac{d}{dt}A_{n}=A_{n+1}-2A_{n}+A_{n-1} .
\end{equation}
For wavelengths much longer than the dimensions of a single SQUID the dispersion relation becomes to lowest order in $k$:
\begin{equation}\label{eq:dispersion}
\omega^{2}(k)=\frac{4}{LC_{0}}\sin^{2}\left(\frac{ka}{2}\right)\approx c^{2}k^{2},
\end{equation}
where we have defined the velocity of propagation as $c=a/\sqrt{LC_{0}}$, which in practice is well below the vacuum speed of light $c_{0}$.  In this limit, the wave equation approaches the continuum
\begin{equation}
\left(\frac{\partial}{\partial t}\frac{1}{c^{2}}\frac{\partial}{\partial t}-\frac{\partial^{2}}{\partial x^{2}}\right)A=0.
\end{equation}
By ignoring higher-order terms in Eq.~(\ref{eq:dispersion}), we effectively remove the discreteness of the array which, along with dispersion from JJ inertia terms, can play the role of Planck scale physics in our system \cite{unruh:2005,philbin:2008,jacobson:1991}.  For parameter values considered below, the relevant short distance scale is $c/\omega_{p}~(>a)$.  Requiring the propagation speed to vary in \textit{both} space and time,
\begin{equation}\label{eq:c}
c^{2}\rightarrow c^{2}(x-ut),
\end{equation}
with fixed velocity $u$ set by an external flux bias pulse, the wave equation in the comoving frame becomes
\begin{equation}\label{eq:wave}
\left[\left(\frac{\partial}{\partial t}-u\frac{\partial}{\partial x}\right)\frac{1}{c^{2}}\left(\frac{\partial}{\partial t}-u\frac{\partial}{\partial x}\right)-\frac{\partial^{2}}{\partial x^{2}}\right]A=0,
\end{equation}
where $x$ and $t$ now label the comoving coordinates.  This wave equation can be re-expressed in terms of an effective space-time metric,
\begin{equation}\label{eq:metric}
g_{\mathrm{eff}}^{\mu\nu}=\frac{1}{c^{2}}\begin{pmatrix}
 1 & -u  \\ 
-u & u^{2}-c^{2}
\end{pmatrix}.
\end{equation}
Comparing this metric with Eq.~(\ref{eq:painleve}), we see that our system contains a horizon located wherever $u^{2}=c^{2}(x)$.  In Fig.~\ref{fig:regions} we plot the effect of a step-like hyperbolic tangent flux bias pulse of amplitude $\Phi_{\mathrm{ext}}=0.2\Phi_{0}$ on a JJA with inductances given by Eq.~(\ref{eq:inductance}), where we have kept only the lowest term in the $I_{c}/I^{s}_{c}$ expansion.
Additionally, since $\Phi_{\mathrm{ext}}$ can only increase the inductance, the flux-bias pulse velocity $u$ must be below the unbiased transmission line propagation velocity $c$ in order to establish a horizon.  We do not consider Gaussian or similar pulse shapes as they generate both black hole and white hole horizons \cite{hawking:1976} which complicates interpretation of the emission process.

So far, we have focused on demonstrating a classical effective background geometry with an event horizon. The next step is to quantize small perturbations in the potential field $A$ about this background. The correct commutation relations between quantum field operators are required for conversion of vacuum fluctuations into photons \cite{unruh:2003}.  These relations have been verified in the systems to which ours is analogous \cite{schutzhold:2005}.  The resulting Hawking temperature  is determined by the gradient of the JJA velocity at the horizon
\begin{equation}\label{eq:temp}
T_{H}=\frac{\hbar}{2\pi k_{b}}\left|\frac{\partial c(x)}{\partial x}\right|_{c^{2}=u^{2}}.
\end{equation}
The radiated power in the comoving frame coincides with the optimal rate for single-channel bosonic heat flow in one-dimension \cite{schutzhold:2005,blencowe:2000,meschke:2006}
\begin{equation}\label{eq:power}
\frac{dE}{dt}=\frac{\pi}{12\hbar}\left(k_{b}T_{H}\right)^{2}.
\end{equation}
Eq.~(\ref{eq:power}) is universal for bosons since the channel-dependent group velocity and density of states cancel each other in one dimension \cite{blencowe:2000}. For a detector at the end of the transmission line, the radiation emitted by an incoming bias pulse will be doppler shifted yielding higher power compared to Eq.~(\ref{eq:power}).  However, the rate of emitted photons remains approximately unchanged.
 
\begin{figure}[htbp]
\begin{center}
\includegraphics[width=3.1in]{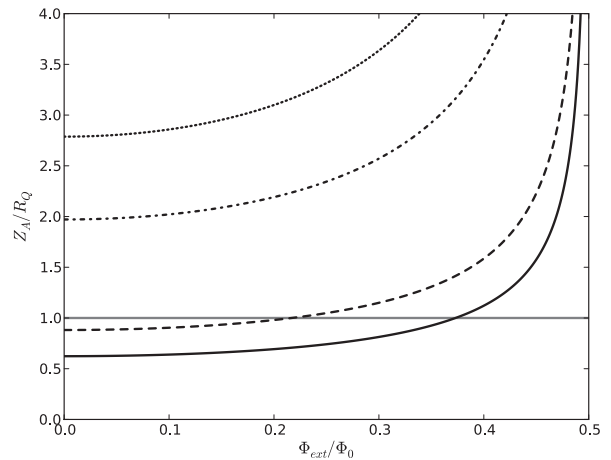}
\caption{Ratio of array impedance $Z_{A}$ to the resistance quantum $R_{Q}$ as a function of the external flux bias for a critical current $I_{c}=2~\mathrm{\mu A}$ and example ground capacitor values: $C_{0}$= $10^{-16}$(solid), $5\times10^{-17}$(dashed), $10^{-17}$(dash-dot), and $5\times10^{-18}\mathrm{F}$(dotted).  High and low impedance regions are defined above and below $Z_{A}/R_{Q}=1$ respectively.}
\label{fig:impedance}
\end{center}
\end{figure}

\textit{Model Validity}.--- For a single effective JJ, the magnitude of quantum fluctuations in the phase variable $\gamma_{+}$ depends on both the ratio of Josephson energy, $E_{J}=\Phi_{0}I^{s}_{c}/2\pi$, to charging energy, $E_{C}=e^{2}/4C_{J}$, as well as on the impedance of the junction's electromagnetic environment.  These energy scales give a representation of the phase-charge uncertainty relation $\Delta\gamma\Delta Q\ge e$, and relate the amplitude of quantum fluctuations between these variables \cite{likharev:1986}.   When $E_{J}/E_{C}>1$ and the impedance seen by the junction is less than the resistance quantum, $R_{Q}=h/4e^{2}\approx6.45\mathrm{k}\Omega$, the phase operator behaves as a semiclassical quantity, i.e. the $\gamma_{+}$ quantum fluctuations are small with respect to its average, and the JJ is in the superconducting state, allowing for a lumped inductor approximation.  In the majority of experimental configurations, a single JJ is connected to probe leads with impedance $\sim50\Omega$ and as such is in the low-impedance regime $Z/R_{Q}\ll 1$.  In contrast, a JJA has an environment that comprises not only the input and output ports, but also all the other JJ's in the array.  In this case, we can define an effective impedance as seen by a single junction to be $Z_{J}=Z_{E}+Z_{A}$ where $Z_{E}$ is the environmental impedance of the leads and $Z_{A}$ is the array impedance that, for frequencies below the plasma frequency, can be written as \cite{haviland:2000}
\begin{equation}\label{eq:impedance}
Z_{A}=R_{Q}\sqrt{\frac{4E_{C}}{E_{J}}}\sqrt{\frac{C_{J}}{C_{0}}}=R_{Q}\sqrt{\frac{2\pi e^{2}}{\Phi_{0}C_{0}I_{c}}\sec\left(\pi\Phi_{\mathrm{ext}}/\Phi_{0}\right)},
\end{equation}
where the last equality explicitly shows the dependence on the external flux bias and single junction parameters.  Thus, even for a small energy ratio $E_{C}/E_{J}$, the lumped inductor model applies only when $Z_{A}/R_{Q}\lesssim 1$.   In Fig.~\ref{fig:impedance} we show the dependence of array impedance $Z_{A}$ on the external bias for fixed critical current $I_{c}=2~\mu\mathrm{A}$ and a range of experimentally valid capacitances to ground.
As $\Phi_{\mathrm{ext}}\rightarrow \Phi_{0}/2$, high impedance causes large phase fluctuations, indicating a breakdown of our semiclassical description; the array undergoes a quantum phase transition from superconducting to insulating Coulomb blockade behavior \cite{chow:1998}.  Note, the small JJ parameter variability in actual arrays \cite{beltran:2007,beltran:2008,chow:1998} will prevent the divergence in Fig.~\ref{fig:impedance}, as well as cause some transmission line scattering in the low impedance superconducting state.

The dependence of the Josephson energy $E_{J}$ on external flux $\Phi_{\mathrm{ext}}$ as described above allows for the systematic introduction of quantum fluctuations in our model.  With the phase variable governing the circuit inductance, these fluctuations manifest themselves in the effective metric (\ref{eq:metric}) through the propagation velocity $c$. As the amplitude of fluctuations increases, the metric becomes a quantum dynamical variable which must be included in the description of the Hawking process.  Thus, consequences of back-reaction from the Hawking process as well as quantum dynamical space-time can be probed by this configuration.  Both processes, not included in the original Hawking derivation, represent analogue quantum gravitational effects present in our system \cite{thiemann:2007}.

\begin{figure}[htbp]
\begin{center}
\includegraphics[width=3.2in]{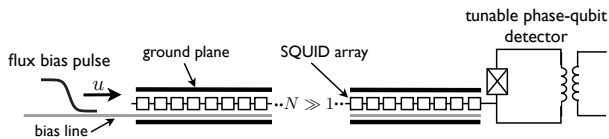}
\caption{Possible transmission line and detector realization.  Current pulses in the bias-line provide external flux necessary to modify the SQUID array propagation velocity.  A phase-qubit at the end of the JJA functions as the photon detector.}
\label{fig:experiment}
\end{center}
\end{figure}

\textit{Experimental Realization}.--- A possible realization of the JJA is shown in Fig.~\ref{fig:experiment}, which consists of the JJA transmission line as well as an additional conducting line producing the space-time varying external flux bias $\Phi_{\mathrm{ext}}$.  To provide a space-time changing velocity, the JJA is modulated by generating current pulses in the bias-line, the propagation velocity of which are assumed to be slightly below that of the unbiased JJA.  The required bias pulse velocities $u$ can be achieved by similarly employing individual JJ's in series as the bias line.  Additionally, a dc-external flux can be used to fine-tune the transmission line velocity closer to that of the bias-line, eliminating the need for large amplitude current bias pulses.  Unavoidable current pulse dispersion in the bias-line, resulting in a decrease in Hawking temperature, can be minimized with appropriate choice of pulse shape and transmission line length.

Unambiguous verification of the Hawking process will require frequency-tunable, single-shot photon detection at the  end of the JJA opposite to that of the bias pulse origin.  Although not presently available, microwave single-photon detectors based on superconducting qubits are under active investigation \cite{wang:2008,romero:2009}.  We will assume a phase-qubit as our model detector \cite{wang:2008}.  By repeatedly injecting current pulses down the bias-line, the predicted blackbody spectrum associated with the Hawking process can be probed by tuning the qubit resonant frequency.  Correlations across the horizon between the emitted photon pairs can be established through coincidence detection.  We emphasize the essential need for correlation information in order to establish that a photon is produced by the Hawking effect rather than some other ambient emission process, or spuriously generated via capacitive coupling to the bias-line.  Unwanted directional coupling can be minimized with proper engineering of the transmission line.

To estimate the Hawking temperature we will assume parameters similar to those of Ref.~\cite{beltran:2007}, with SQUID's composed of tunnel junctions with $I_{c}=2~\mathrm{\mu A}$ and an upper bound achievable plasma frequency $\omega_{p}=2\pi\times10^{12}~\mathrm{Hz}$.  The capacitance to ground is assumed to be $C_{0}=5\times10^{-17}~\mathrm{F}$ (dashed line in Fig.~\ref{fig:impedance}).  Using a SQUID length $a=0.25~\mu\mathrm{m}$ gives an unbiased transmission line velocity $c\sim c_{0}/100$.  Equation~(\ref{eq:temp}) gives the temperature as determined by the rate at which the JJA transmission line velocity varies that, in our case, is limited by the plasma frequency $\omega^{s}_{p}$.  Assuming the maximum rate is an order of magnitude below $\omega^{s}_{p}\left(\Phi_{\mathrm{ext}}=0\right)/2\pi$, then the Hawking temperature is $\sim 120~\mathrm{mK}$.  This temperature can be a factor of ten larger than the ambient temperature set by a dilution refrigerator and therefore should be visible above the background thermal spectrum.  Using Eq.~(\ref{eq:power}) and the sample pulse in Fig.~(\ref{fig:regions}) gives an initial Hawking temperature $120~\mathrm{mK}$, which decreases $\sim10\%$ every $1000$ JJA elements due to bias-line dispersion.  Applying the power expression (\ref{eq:power}) yields an average emission rate of one photon per pulse for $\sim 4800$ SQUID's. Of course, the transmission line can be made considerably shorter at the expense of an increase in number of pulse repetitions in order to accumulate sufficient photon counts to verify the Hawking radiation.  The parameters and pulse shapes chosen here illustrate feasibility of this setup, but do not represent the only available configuration.  These values can likely be improved upon and optimized in terms of both performance and fabrication of this proposal. 

\textit{Conclusion}.--- We have demonstrated that an array of dc-SQUID's in a coplanar transmission line, when biased by a space-time dependent flux, creates an effective space-time metric with a horizon.  As a quantum device, the superconducting transmission line allows for the possibility of observing not only the Hawking effect, but also the effects of quantum fluctuations in an analogue gravitational system.

This work was supported by the U.S.-Israel Binational Science Foundation (BSF) and the National Science Foundation (NSF) under Grant Nos. DMR-0804477 and DMR-0804488 .

\bibliography{analogueBIB}
\end{document}